\documentclass[twocolumn,preprintnumbers,amsmath,amssymb,superscriptaddress,prl]{revtex4}

\usepackage[T1]{fontenc}
\usepackage{lmodern}
\usepackage[utf8]{inputenc}
\usepackage[english]{babel}
\usepackage{graphicx}
\usepackage{dcolumn}
\usepackage{bm}
\usepackage{color}
\usepackage{soul}

\begin{document}

\title{The morphology transition mechanism from icosahedral to decahedral phase during growth of nanoclusters}

\author{Alexey A. Tal}
\email{aleta@ifm.liu.se}
\affiliation{Theory and Modeling, IFM-Material Physics, Linköping University, SE-581 83, Linköping, Sweden}
\affiliation{Materials Modeling and Development Laboratory, National University of Science and Technology 'MISIS', 119049, Moscow, Russia}  
\author{E. Peter Müger}%
\affiliation{Theory and Modeling, IFM-Material Physics, Linköping University, SE-581 83, Linköping, Sweden}%

\author{Igor A. Abrikosov}
\affiliation{Theory and Modeling, IFM-Material Physics, Linköping University, SE-581 83, Linköping, Sweden}%
\affiliation{Materials Modeling and Development Laboratory, National University of Science and Technology 'MISIS', 119049, Moscow, Russia}  

\begin{abstract}
The morphology transition from the thermodynamically favorable to the unfavorable phase during growth of free-standing copper nanoclusters is studied by molecular dynamics simulations. We give a detailed description of the kinetics and thermodynamics of the process. A universal mechanism of a solid-solid transition, from icosahedral to decahedral morphology in nanoclusters, is proposed. We show that  a formation of distorted NC during the growth process with islands of incoming atoms localized in certain parts of the grown particle may shift the energy balance between Ih and Dh phases in favour of the latter leading to the morphology transition deep within the thermodynamic stability field of the former. The role of diffusion in the morphology transition is revealed. In particular, it is shown that fast diffusion should suppress the morphology transition and favour homogeneous growth of the nanoclusters. 

\end{abstract}

\maketitle

The study of nanoclusters (NC) growth has recently been in focus of many intense research activities \cite{baletto05}, because of their unique properties and also the fundamentally new physical effects that occur in finite-sized systems \cite{ekimov82}. NCs have found broad applications in catalysis \cite{prieto13,cuenya10,henry98} biomedicine or photovoltaic \cite{taton00,chao02,garcia11}. All the applications require a precise control of the NC growth process and understanding of their properties. One of the most significant properties of the NC is their morphology. It was shown that the favorable structure for copper nanocluster for N<1000 atoms is the Mackay icosahedron (Ih) \cite{uppenbrink91,wales96}, followed by the Marks decahedron (Dh)\cite{marks85}, which corresponds to the minimum energy structure for 1000<N<30000 atoms in the NC. At larger N the clusters optimal structure is FCC \cite{baletto02}. However, experiments demonstrate a significant amount of energetically unfavorable morphologies for different synthesis methods \cite{koga04, feng14}. It is commonly accepted that it happens due to kinetics of the growth process. To study the growth kinetics is thus of fundamental importance. Baletto et al. \cite{baletto00} have demonstrated a possibility of the morphology transition in molecular dynamics simulations of silver nanoclusters and explained its mechanism \cite{baletto01}.  They claimed that the transition from icosahedral (Ih) to decahedral (Dh) morphology goes through a melting of the cluster and a formation of the new morphology from the melted cluster. Lan et al. \cite{lan14} proposed a qualitatively different mechanism for the transformation from Dh to Ih, where morphology goes through a solid-solid transition without the formation of an amorphous phase. On the other hand, it is known that the smaller the cluster size is the more favorable the Ih phase should be. This means that in a real growth process the transition should go from Ih to Dh and it is very unlikely to go backwards. Thus it is much more relevant to study the Ih-Dh transition. 

In this Letter, we report the mechanism of the solid-solid morphology transition that we have discovered in copper nanoclusters. Also, we provide a detailed explanation of the conditions that induce the morphology transition, or on the contrary preserve the layer-by-layer growth. Also, the decisive role of diffusion in the morphology transition is exhaustively explained. 

Note that the main interest in MD simulations of NCs addressed low-energy nanocluster growth techniques. However, recent developments of new NC synthesis methods, e.g. employing pulsed highly ionized plasma \cite{pilch13a,pilch13b}, have drawn attention to higher energies during the growth. The use of a plasma environment has several important advantages such as an increased growth rate and a wide choice of possible materials to grow. It has already been shown that the energy of the growth process significantly affects the kinetics of the growth process \cite{tal14}, a fact that should be accounted for in simulations. For instance, it was shown that the Coulomb interaction between the NC and impinging ions during the growth process will influence the angular distribution of the velocities of the impinging atoms. Consequently the angular distribution will affect the diffusion and local heating of the cluster surface.

We used the embedded-atom method (EAM) potential with Foiles parametrization \cite{foiles86} to simulate the growth of copper NCs. We start our growth simulations from a seed. Based on the Baletto's analysis of structures and magic numbers for clusters \cite{baletto02}, an icosahedral seed consisting of 147 atoms was chosen as our growth seed.  In the simulation new atoms were randomly generated every 100~ps around the cluster on a sphere with radius 13 Å. That is, slightly outside the cutoff radius of the potential from the NC.  Though the growth rates in our simulations are much higher than the experimental ones,  we employ a thermostat to cool the cluster down to the correct temperature before a new particle impinge, this provides the correct conditions for the growth. The final structure was analyzed after that 300 atoms were added to the seed. The temperature of the nanocluster was controlled by a Nosé-Hoover thermostat \cite{nose84} and the speed of the impinging atoms corresponded to either the temperature of the cluster or a kinetic energy of 1~eV. These two cases were chosen to represent the inert-gas-aggregation (IGA) growth process or the conditions of the growth in a plasma, respectively  \cite{pilch13a}.

We performed simulations in LAMMPS \cite{lammps} for different temperatures, incident velocities and angular distribution of the impinging atoms \cite{tal14}. Every considered temperature, velocity and angular distribution of the impinging atoms were studied in series of 50 simulations. Analysis of the results showed significant fractions (up to 30\% ) of non-icosahedral clusters among the final clusters. All results are assembled in Table I. Normal incidence and full distribution correspond to particles grown from atoms with an energy of 1~eV, simulating the NC growth in plasma. We have considered two models of the growth, with atoms impinging on the surface with normal incidence and with the angular distribution derived in \cite{tal14}. Results denoted as IGA  correspond to a respective process simulated with the assumption that the impinging particles have a thermal energy (0.03~eV). Note that, as it was mentioned early, the thermodynamically favorable morphology for clusters of this size is icosahedral. Thus the kinetics of the growth process induces the morphology transition. However, the majority of the final clusters are still icosahedral, which rises the question: What is the difference between these two outcomes?

To address this question, we analyze in details one typical simulation, where the structure changes from icosahedral to decahedral during the growth. Fig.1 shows coordination of atoms calculated with Bond-angle analysis (BAA) \cite{stukowski12}. The method was developed by Ackland and Jones \cite{ackland06} to distinguish fcc, hcp and bcc coordination structures. From the bond vectors of the central atom the histogram of the bond angle cosines is computed and then used to determine the structure type by the heuristic decision rules. Initially the cluster consists of 147 atoms arranged in a perfect icosahedron with all atoms HCP-coordinated. However, all surface atoms are identified as disordered by the BAA method. Thus we see that $\approx$ 50 interior atoms are HCP-coordinated and $\approx$ 100 surface atoms are disordered. At the size of $\approx$ 250 atoms the cluster undergoes the morphology transition from icosahedral to decahedral. At that point we see how the number of HCP-coordinated atoms drops down and the number of FCC-coordinated atoms abruptly increases. It clearly shows that part of the HCP-coordinated atoms change their local structure to FCC-coordination, since we don't see any significant change in disordered atoms. After the transition FCC-coordinated atoms dominate over the HCP-coordinated ones for the remaining of the growth simulation. This means that there were no more morphology transition.  

\begin{figure}[h]
\begin{center}
\includegraphics[width=0.5\textwidth]{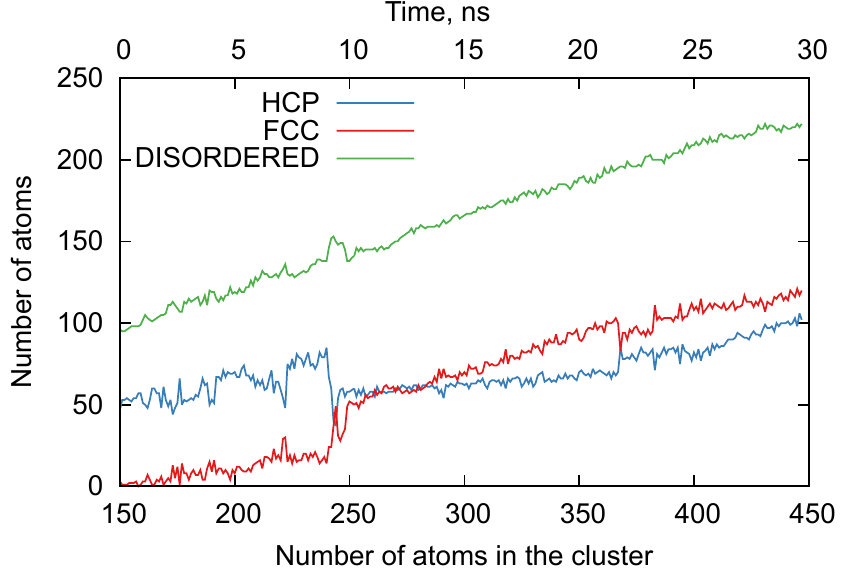}
\end{center}
\caption{Bond-angle analysis of the cluster structure. The figure shows how the number of HCP, FCC and disordered atoms changes with time, or equivalently the cluster size, during the growth process.}
\end{figure}

In order to understand what induces the transition we analyzed the potential energy of the cluster as a function of time. Fig.2 shows the typical case described in the proceeding paragraph where the morphology transition takes place between incident atom 101 and 102, i.e. when the cluster has 248 atoms. We see that from 20 till 35 ps, after atom 101 were introduced, the potential energy of the cluster fluctuates. This means that the structure is far from perfect and that atoms on the surface change their positions and thereby change the potential energy of the system. However, after the transition, the structure falls into a potential well, where it minimizes the potential energy and thus we see no more fluctuations of the energy. It is worth mentioning that the full transition takes approximately 2~ps and the whole time range of Fig.2 corresponds to one point in Fig.1. The energy between these two states is 1.8~eV. Also one can see in Fig.2 that the surface area of the cluster changes significantly at the transition. The fluctuations of the potential energy correspond to changes of the surface area until the clusters undergoes the morphology transition. Then the surface area promptly increases while fluctuations vanish.

We applied the Nudged Elastic Band (NEB) method to obtain the height of the potential energy barrier for the transition from icosahedral to decahedral morphology, Fig.2a. This allows us to plot a Potential Energy Surface (PES) of this particular cooper nanocluster consisting of 248 atoms, Fig.3. The icosahedral phase corresponds to a minimum with a potential energy of -790.5~eV and the decahedral phase corresponds to a deeper local minimum with a potential energy of -792.3~eV. The barrier the cluster needs to overcome in order to minimize its energy is 0.8~eV. Besides, we calculated the barrier heights for different sizes of NCs. The cluster consisting of 185 atoms has a 0.6~eV barrier and the 376-atom cluster has to overcome a barrier of 2~eV.  

\begin{figure}[]
\begin{center}
\includegraphics[width=0.5\textwidth]{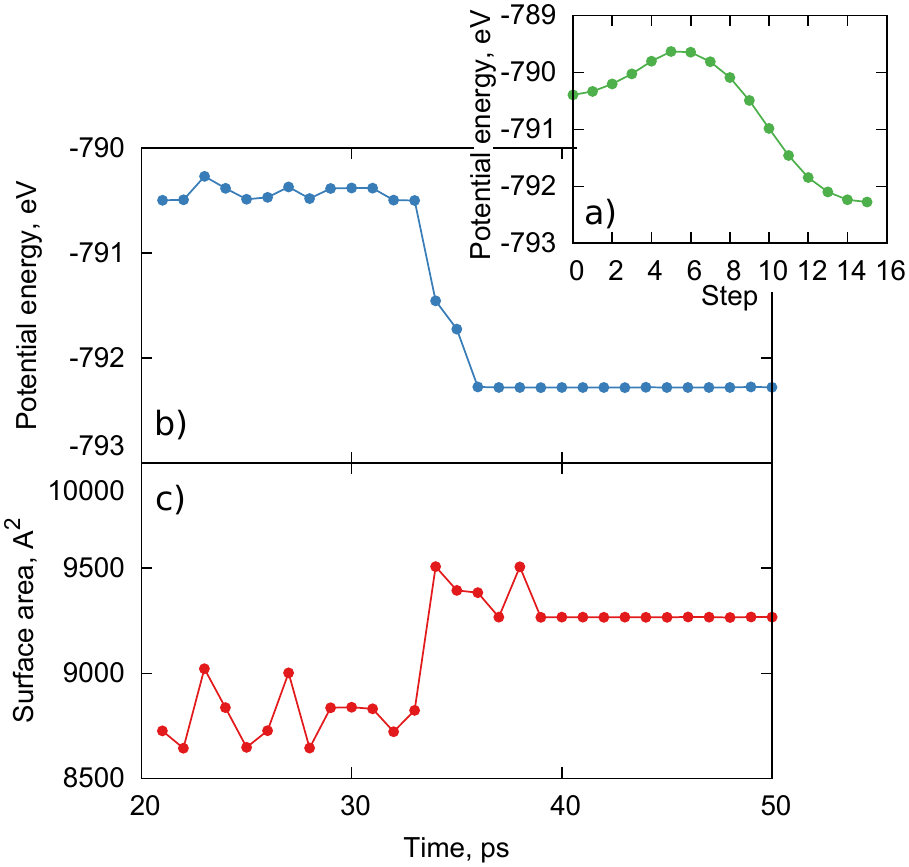}
\end{center}
\caption{Data from the same simulation as in Fig.1 but only for a short time period when the cluster consist of 248 atoms. (a) NEB analysis of the transition from the Ih to the Dh phase; (b) potential energy of the cluster as a function of time; (c) dependence of the cluster surface area on the simulation time.}
\end{figure}

\begin{figure}[]
\begin{center}
\includegraphics[width=0.5\textwidth]{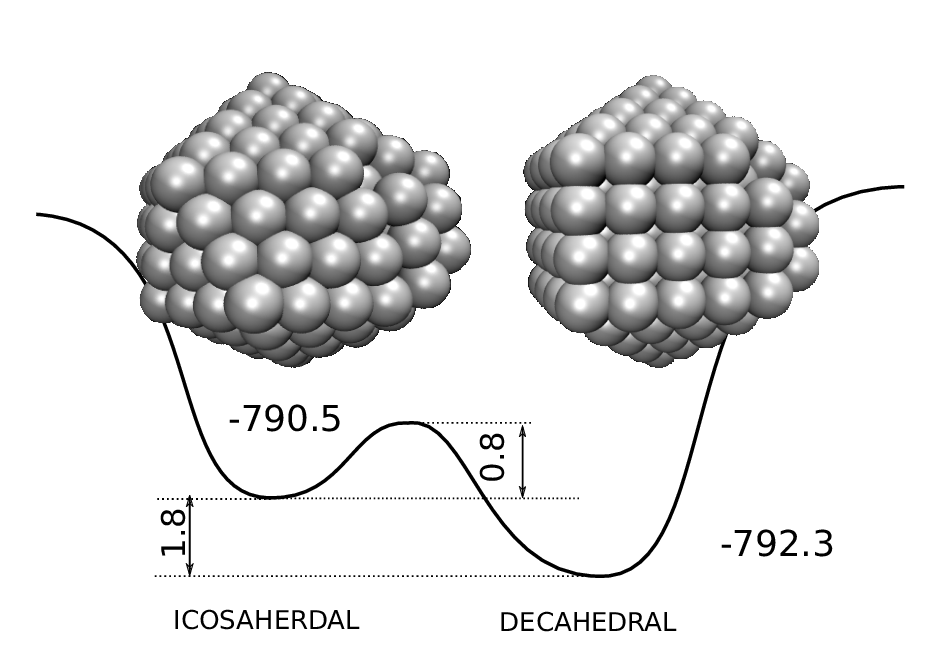}
\end{center}
\caption{The potential energy surface for the same 248 atom copper nanocluster as in Fig.1 and Fig.2}
\end{figure}

\begin{figure*}[]
\begin{center}
\includegraphics[width=1\textwidth]{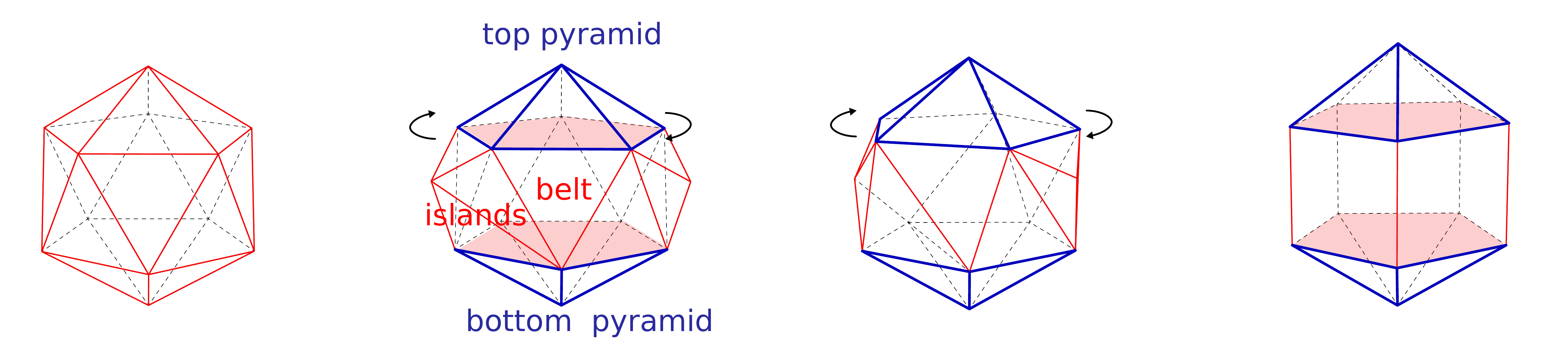}
\end{center}
\caption{Mechanism of the morphology transitions from Ih to Dh phase: a) A perfect icosahedron; b) impinging atoms form islands on the surface of the cluster, thus increasing the surface energy, the most disturbed facets could be described as located in some kind of "belt"; c) the bottom and top pentagonal pyramids twist with respect to each other and a fractional part of the islands on the facets incorporates into the cluster layers; d) aligned edges of the new {100} facets and with all atoms in between the slices rearrange into a FCC structure. }
\end{figure*}

It is known, that icosahedron is the favorable morphology in many noble metals \cite{baletto02} at small sizes due to its small surface area (see Fig.4a). Thus icosahedron morphology minimizes the surface area, but causes relatively large internal stress. On the contrary, the decahedron morphology has a smaller internal stress but a larger surface area. If the impinging atoms during the growth are not able to diffuse far enough to form an uniform new layer, and thus instead form islands on the facets of the icosahderon Fig.4b, then the surface area significantly increases. Thus, the surface area of the non-perfect icosahedron is large as well as its internal stress. A transition to the decahedral phase minimizes the internal stress though the surface area increases. Still, as one sees in Fig. 3a, the potential energy of the decahedral phase is lower than that of the imperfect icosahedron.  
\begin{center}
\begin{table}{}  
\caption{The fraction of the clusters that suffered a morphology transition for different temperatures. Normal incidence (NI) correspond to particles grown from atoms impinging on the cluster with normal incidence and a kinetic energy of 1~eV. Full distribution (FD) corresponds to particles grown from atoms with kinetic energy of 1~eV and a distributions of incidences from normal to grazing. Particles grown from atoms with a kinetic energy of 0.03~eV represent the IGA process.}
 
  \begin{tabular}{ c    c   c c  }
\hline \hline    
    Temp, K &  NI, \% & FD, \% & IGA,\%\\
\hline   
    400 & 28 & 16 & 26  \\ 
    450 & 20 & 16 & 22\\ 
    500 & 10 & 10 & 18\\ 
    600 & 8 & 8 & 12\\ 
    700 & 4 & 2 & 2\\ 
 \hline  
 \hline 
  \end{tabular}
\end{table}
\end{center}
Another way to minimize the energy is to redistribute the impinging atoms more uniformly, thus preventing islands to form on the cluster surface and strong expansion of the surface area. Statistics of the simulated clusters  (Table I) witnesses that the higher the temperature the lower the probability of the transition to the decahedral phase. This fact can be explained by the rate of diffusion, which is higher at high temperatures. Moreover we have observed that small islands may diffuse, thus redistributing extra atoms more uniformly on the surface of the cluster and thereby decreasing the surface area. At the same time, it is clear that growth processes that restricts diffusion like growth with normal incidence particles or the IGA process, where the diffusion is limited by the low energy of the impinging particle, have noticeably higher statistics for the morphology transitions. All these observations demonstrate how important the diffusion is in the morphology transition. But exactly how does a non-perfect icosahedron transform into a decahedron? 

Detailed observations of hundreds of our NC growth simulations allowed us to identify a pattern and describe the mechanism of the morphology transitions from Ih to Dh phase. This mechanism has very simple foundations and thus seems very elegant. The basic idea is the following: the formation of islands on the surface disturbs the perfect facets of the icosahderon (Fig. 4a). This disturbance can be localized in a certain area of the NC, which could be considered as a "belt" formed by the extra atoms, Fig4b. Above and under this "belt" one can distinguish two pentagonal pyramids with much less disorder. These pyramids do not require much rearrangement, because they should remain the same in the decahedral cluster. They only need to be twisted with respect to each other in order to align their edges and form the new rectangular facets of a decahedron Fig4c-d. During these relative rotations the atoms in the "belt" undergo major rearrangements. This includes the formation of FCC-coordinated atoms from HCP ones and incorporation of the extra atoms from the surface into new layers. 

Baletto et al. \cite{baletto00} suggested that this transition in silver nanoclusters goes through a melted phase of the cluster and that the new phase is formed from an amorphous structure. However, Lan et al. \cite{lan14} gave very serious arguments in favor of a solid-solid transition. On the other hand, Lan et al. showed a transition in "perfect" clusters (cluster with magic number of atoms) from the least favorable to the most favorable phase in a transition that has been driven by pure thermodynamics. Thus they did not account for the influence of the kinetics, which is believed to be the morphology-determining process at realistic conditions. Koga et al. \cite{koga04} suggested a 
mechanism similar to ours from experiments of gold nanoclusters growth. We have analyzed more than 700 simulations and we have shown that the mechanism of the morphology transition remains the same for all considered sizes, temperatures and even energies of the impinging particles. This allows us to claim that this is the universal mechanism of the solid-solid Ih-Dh morphology transition for all copper nanoclusters.  That gives us a very rigid reason to expect it to be the universal mechanism for all metal nanoclusters. 

In summary, we have reported the mechanism of solid-solid transition form Ih to Dh phase in metal nanoclusters, and suggested an explanation of the cause of a morphology transition in Ih nanoclusters deep within their thermodynamic stability field. The origin of the transition has been associated with a formation of distorted NC during the growth process with islands of incoming atoms localized in certain parts of the grown cluster. The distortions change the energy balance between Ih and Dh phases, leading to the changes of the morphology of the NC. We revealed the role of diffusion on the morphology transition and showed how different facets of the cluster will transform during the transition. The fundamental understanding of the morphology transition mechanism allows us to suggest that varying the diffusion length in the growth process can be used to influence the cluster morphology. Considering the fact that the transition barrier increases with the increase of the cluster size we expect that once the morphology is set at a certain large size it is very unlikely to change. A fact that again seems very convenient for tailoring the morphology of the grown nanoclusters.  

The work was financially supported by the Knut and Alice Wallenberg Foundation through Grant No. 2012.0083. I.A.A. is grateful for the support provided by the Swedish Foundation for Strategic Research (SSF) program SRL Grant No. 10-0026. The support by the Grant from the Ministry of Education and Science of the Russian Federation (Grant No. 14.Y26.31.0005) is gratefully acknowledged. The calculations were performed on resources provided by the Swedish National Infrastructure for Computing (SNIC) at the National Supercomputer Center (NSC).

\bibliographystyle{plain}

\end{document}